\documentclass[12pt]{article}
\usepackage[dvipdfmx]{hyperref}
\usepackage{amssymb,amsmath,graphicx}

\newcommand{\LQ}{\Lambda_{\rm QCD}}

\newcommand{\msbar}{$\overline{\rm MS}$}
\newcommand{\mbar}{\overline{m}}

\begin{document}

\begin{titlepage}

    \begin{flushright}
      \normalsize TU--1085\\
      \today
    \end{flushright}

\vskip2.5cm
\begin{center}
\Large
UV contributions to energy of a static quark-antiquark pair\\
in large-$\beta_0$ approximation
%\unboldmath
\end{center}

\vspace*{0.8cm}
\begin{center}
{\sc Yuuki Hayashi} and
{\sc Yukinari Sumino}\\[5mm]
  {\small\it Department of Physics, Tohoku University}\\[0.1cm]
  {\small\it Sendai, 980-8578 Japan}

\end{center}

\vspace*{2.8cm}
\begin{abstract}
\small
%\noindent
The total energy
of a static quark-antiquark pair 
$E_{\rm tot}(r)=2m_{\rm pole}+V_{\rm QCD}(r)$
is known to 
include
%be predictable up to 
${\cal O}(\LQ^3 r^2)$ and ${\cal O}(\LQ^3/\overline{m}^2)$ renormalon uncertainties
in the large-$\beta_0$ approximation,
after canceling ${\cal O}(\LQ)$ renormalons.
%We compute the predictable part (genuine UV part) 
We compute the (renormalon-free) genuine UV part 
in terms of the {\msbar} mass
$\overline{m}$, extending a recently-proposed method
which conforms with OPE.
In particular the $r$-independent part is determined.
The result would help understanding the nature of $E_{\rm tot}(r)$
in the context of OPE with renormalon subtraction.
\vspace*{0.8cm}
\noindent
%PACS numbers: 

\end{abstract}

% {\it Email addresses}:

% \texttt{anzai@tuhep.phys.tohoku.ac.jp},
% \texttt{sumino@tuhep.phys.tohoku.ac.jp}.
%\hspace{36.5mm}
\vfil
\end{titlepage}

\newpage

%\section{Introduction}
%\label{}

Being much smaller than ordinary hadrons,
a hadron composed of a heavy quark and its anti-particle (heavy quarkonium)
is an ideal system
that can be systematically analyzed using solid analysis tools of the strong
interaction, such as
perturbative QCD and operator
product expansion (OPE).
In particular various analyses of the leading-order total energy of this system, 
defined by
$E_{\rm tot}(r)\equiv 2m_{\rm pole}+V_{\rm QCD}(r)$,
have provided a lot of insight into the theoretical structure of perturbative QCD and 
OPE.
The analyses were brought to a new phase
by the discovery of the cancellation of ${\cal O}(\LQ)$
renormalons in $E_{\rm tot}(r)$ 
\cite{Pineda:id,Hoang:1998nz,Beneke:1998rk}, 
which led to a dramatic improvement in
convergence of perturbative series of $E_{\rm tot}(r)$
and to much more accurate computation.

Series of studies 
\cite{Sumino:2003yp,Sumino:2004ht,Sumino:2005cq,Mishima:2016xuj,Mishima:2016vna} 
on the static QCD potential
$V_{\rm QCD}(r)$ in perturbative QCD have shown that the potential
can be expressed in an expansion in $r$ in the form
\begin{align}
V_{\rm QCD}(r)=V_C(r) + C_0^V + C_1^V  r + {\cal O}(\LQ^3 r^2) 
~~~\mbox{for}~~r\ll \LQ^{-1}
\label{expVQCD}
\, ,
\end{align}
by resumming logarithms by
renormalization group (RG) or in the large-$\beta_0$ approximation.
Here, the $r$-independent constant $C_0^V$ includes ${\cal O}(\LQ)$
renormalon;
$V_C(r)$ and $C_1^V  r$ correspond to genuinely ultraviolet (UV) part and
can be computed without renormalon 
uncertainties.
$V_C(r)$ has a Coulomb-like form with logarithmic corrections
at short-distances.\footnote{
The logarithmic corrections render the short-distance
behavior of $V_{\rm QCD}(r)$ 
to be consistent with the RG equation.
At large $r$, $V_C(r)$ approaches a pure Coulomb potential.
}
The above expansion in $r$ is consistent
with OPE performed in an effective field theory 
(EFT) 
``potential
non-relativistic QCD''
(pNRQCD) \cite{Brambilla:2004jw}.
In fact, $V_C(r)+C_1^V  r$ can be identified with the leading
Wilson coefficient of OPE of $V_{\rm QCD}(r)$ 
after subtraction of renormalons.
This OPE of $V_{\rm QCD}(r)$ is formulated beyond the large-$\beta_0$ approximation
(i.e., including sub-leading logarithms), and recently it has been
applied to a precise determination of $\alpha_s(M_Z)$ by comparing
the above prediction with lattice computation in an OPE framework,
in which the $ {\cal O}(\LQ) $
and $ {\cal O}(\LQ^3 r^2) $ renormalons are subtracted
\cite{Takaura:2018lpw,Takaura:2018vcy}.

Refs.~\cite{Mishima:2016xuj,Mishima:2016vna} have developed a prescription
[referred to as ``Contour Deformation (CD) prescription'' hereafter], which 
performs
an expansion of 
a general observable $X(Q)$ in the inverse of 
the hard scale $Q$,
after resummation to all orders in $\alpha_s$
within the large-$\beta_0$ approximation. 
($1/Q=r$ in the case of the
QCD potential.)
This prescription achieves separation of UV and infrared (IR) contributions 
in a natural way.
Furthermore, a detailed connection of this expansion
to OPE is given through
the expansion-by-region technique.

Similarly to the QCD potential, we expect that, 
when expressed in terms of a short-distance mass\footnote{
``Short-distance mass'' stands for a class of quark mass definitions
which contain only contributions from UV degrees of freedom to
the quark self-energy in the renormalization.
See e.g.\ \cite{Corcella:2019tgt,Kiyo:2015ooa}
for various definitions and their comparisons.
} 
$\overline{m}$,  
the pole mass of a heavy quark can
be expressed in the form
\begin{align}
m_{\rm pole}=\overline{m} + \Delta M_0(\overline{m}) + C_0^m 
+  \frac{C_1^m}{\overline{m}}  + {\cal O}(\LQ^3/ \overline{m}^2) 
~~~\mbox{for}~~\mbar \gg \LQ
\label{expmpole}
\, ,
\end{align}
where the $\overline{m}$-independent constant $C_0^m$ includes ${\cal O}(\LQ)$
renormalon\footnote{
It is suggested that the ${C_1^m}/{\overline{m}} $ term also includes a
renormalon beyond the large-$\beta_0$ approximation.
We discuss this issue at the end of the paper.
};
$\Delta M_0(\mbar)$ denotes the part proportional to $\mbar$ with
logarithmic corrections at large $\mbar$.
Nevertheless, even in the case
restricting to the large-$\beta_0$ approximation, there is a
difficulty to apply
the CD prescription to carry out this expansion.
The difficulty comes from (UV) renormalization of the pole mass, and
we need to devise an extension of the prescription to deal with it.

As already mentioned, in the combination $2 C_0^m + C_0^V$, the ${\cal O}(\LQ)$
renormalons cancel.
As a result $E_{\rm tot}(r)$ can be computed up to
${\cal O}(\LQ^3 r^2)$ and ${\cal O}(\LQ^3/\overline{m}^2)$ renormalon uncertainties.
The explicit expression of this computable part of $E_{\rm tot}(r)$ should depend on the
definition of the short-distance mass $\mbar$ to be used.
Analytic or semi-analytic analyses of this computable part
have been missing.
The purpose of this paper is to give an explicit expression for it and to
provide a semi-analytic analysis,
in the case we choose
the {\msbar} mass for $\mbar$ and within the large-$\beta_0$ approximation.
An interesting question may be as follows:
Is there a constant term proportional to $\LQ$ (independent of $r$ and $\overline{m}$)
included in this computable part?
[Noting that $C_1^V r =(2 \pi e^{5/3} C_F/\beta_0) \LQ^2 r$ 
in the large-$\beta_0$ approximation, this may not be a
completely absurd question.
For instance, one may suspect a possibility that different short-distance
masses are mutually related by an ${\cal O}(\LQ)$ difference.]

Subtraction of the ${\cal O}(\LQ)$ renormalons from $E_{\rm tot}(r)$
and $m_{\rm pole}$ has also been studied in \cite{Pineda:2001zq,Ayala:2014yxa}
in connection with pNRQCD EFT.
The analyses concern how to subtract the renormalons from
finite-order perturbative series (not restricting to
the large-$\beta_0$ approximation).
A major difference of our method is that (in
the large-$\beta_0$ approximation) we resum the series to all
orders and extract a renormalon-free part in the form
which conforms with OPE, i.e., expansion in $r$ or $1/\mbar$.
In this way we can obtain an insight into the
analytic structure of $E_{\rm tot}(r)$ which would be useful
in the framework of OPE.

In the large-$\beta_0$ approximation, the 
difference of the pole mass and the {\msbar} mass can be 
computed as follows.
\begin{align}
\delta m(\mu) 
&\equiv m_{\rm pole}-\overline{m} (\mu)
\nonumber
\\
&=
-4\pi iC_{F}\overline{m} (\mu)
\int \frac{d^Dq}{\left( 2\pi \right)^D}\frac {2+\left( D-2\right)\dfrac {p\cdot q}
{\overline{m}(\mu)^2}}{\left( q^{2}+2p\cdot q+i0\right) \left( q^{2}+i0\right)} 
\,
\frac{\alpha_s(\mu) \overline{\mu}^{2\epsilon}}{1-\Pi(q^2,\epsilon)}
~+~ ({\rm c.t.}) \, .
\label{deltam}
\end{align}
Here, (c.t.)\ denotes contributions of the diagrams including 
counter terms.
%$\bar{\mu}= {\mu}(e^{\gamma_E}/(4\pi))^{1/2}$.
$\bar{\mu}= \frac{\mu}{\sqrt{4\pi}}\, e^{\gamma_E/2}$,
and $\mu$ represents the renormalization scale in the {\msbar} scheme.
The external momentum is fixed as $p^2=\mbar(\mu)^2$.
We employ dimensional regularization with $D=4-2\epsilon$.
$\Pi(q^2,\epsilon)$ denotes the one-loop vacuum polarization of gluon
in the large-$\beta_0$ approximation (in dimensional regularization).
It is understood that in the end the limit $\epsilon \to 0$ is
taken at each order of the expansion in $\alpha_s(\mu)$.

We can perform Wick rotation and the integral can be brought to a
one-parameter integral 
%with respect to 
over the modulus-squared of the Euclidean
gluon momentum $\tau =q_E^2= -q^2$ \cite{Neubert:1994vb,Beneke:1994qe}
by replacing the
gluon propagator as
\begin{align}
\frac{1}{q_E^2}=\int_0^\infty \frac{d\tau}{\tau}
\, \delta (\tau-q_E^2)
= {\rm Im}\int ^{\infty }_{0}\frac {d\tau }{\pi \tau }\,
\frac{1}{q_E^2-\tau-i0}
\,.
\end{align}
We obtain
\begin{align}
\delta \equiv \frac{\delta m }{\overline{m}}
=
{\rm Im}\int ^{\infty }_{0}\frac {d\tau }{\pi \tau }\,
W_m\!\left(\dfrac{\tau}{\overline{m}^2},\dfrac{\overline{\mu}}{\overline{m}};\epsilon\right)
\,
\frac{\alpha_s(\mu)}{1-\Pi(-\tau,\epsilon)}
~+~ ({\rm c.t.}) \,,
\label{delta-1paramint}
\end{align}
where
\begin{align}
W_m=
\frac { C_{F}}{4\pi }\, \Gamma( \epsilon )\int ^{1}_{0}dx\,
( 2+2x-2\epsilon x) \left( \frac {4\pi\bar{\mu }^{2}}{\overline {m}^{2}x^{2}+\tau x-\tau -i0}\right)^\epsilon  \,.
\end{align}

The difficulty in applying the CD prescription to this integral
representation is as follows.
The prescription, as it is formulated, requires that we can set 
\begin{align}
\frac{\alpha_s(\mu)}{1-\Pi(-\tau,\epsilon)}
~~~\rightarrow~~~
\alpha_{\beta_0}(\tau)=\frac{4\pi/\beta_0}{ \log\left(\tau/\Lambda'^2\right)}
~~~;~~~\Lambda'= e^{5/6}\Lambda_{\rm QCD}
\end{align}
inside the $\tau$ integral.
In eq.~\eqref{delta-1paramint} this is not possible, however, since we
cannot take
the limit $\epsilon\to 0$ before $\tau$ integration due to the UV divergent nature of the integral.
We will circumvent the difficulty by separating $\delta$ into two parts
and introducing a UV cut-off.

We recall the all-order formula of
$\delta $ in $\alpha_s$ expansion \cite{Beneke:1994qe}:
%$\delta $ is known to all orders in $\alpha_s$ expansion:
\begin{align}
\delta 
&=
\frac{C_F\alpha_s(\mu)}{2\pi}\sum_{n=0}^\infty\left(\frac{\beta_0\alpha_s(\mu)}{4\pi}\right)^n\left[G_{n+1}(\mu)\cdot n!+\frac{(-1)^n}{n+1}g_{n+1}\right].
\label{all-order-formula}
\end{align}
The coefficients $G_n$ and $g_n$ are given by 
\begin{align}
&
G(t;\mu)=\sum_{n=0}^\infty G_n(\mu)t^n=\left(\frac{e^{{5}/{3}}\mu^2}{\overline{m}(\mu)^2}\right)^t 3(1-t)\frac{\Gamma(1+t)\Gamma(1-2t)}{\Gamma(3-t)} \,,
\\
&
g(t)= \sum_{n=0}^\infty g_n t^n=\frac{3-2t}{6}\frac{\Gamma(4-2t)}{\Gamma(1+t)\Gamma(2-t)^2\Gamma(3-t)}
\,.
\end{align}
They are related via RG equations to
the anomalous dimension of the running mass and 
the RG-invariant constant terms of ${\delta}$ \cite{Ball:1995ni}:
\begin{align}
&
\gamma_n=\left(\frac{-\beta_0}{4}\right)^{n}C_F g_n
\,,
~~~~~~~~~~
{\rm i.e.}, ~~~
\gamma_m^{(\beta_0)} (\alpha_s)=C_F\, g\!\left(-\frac{\beta_0 \alpha_s}{4\pi}\right)
\,,
\label{gn-anomalousdim}
\\
&
d_n=\frac{C_F}{2}\left(\frac{\beta_0}{4}\right)^n \left[G_{n+1}(\overline{m})n!+\frac{(-1)^n}{n+1}g_{n+1}\right]
~~\mbox{with}~~
\delta 
=
\sum_{n=0}^\infty d_n\left(\frac{\alpha_s(\overline{m})}{\pi}\right)^{n+1}
\,,
\end{align}
where, within the large-$\beta_0$ approximation, the RG equations read
\begin{align}
&
\mu\frac{d}{d\mu}\left(\frac{\alpha_s(\mu)}{\pi}\right)
=-\frac{\beta_0}{2}\left(\frac{\alpha_s(\mu)}{\pi}\right)^2
\,,
\\
&
\mu\frac{d}{d\mu}\overline{m}(\mu)=-\gamma_m^{(\beta_0)} (\alpha_s(\mu)) \overline{m} (\mu)
\,,
~~~~~~
\gamma_m^{(\beta_0)} (\alpha_s(\mu))=\sum_{n=0}^{\infty}\gamma_n\left(\frac{\alpha_s(\mu)}{\pi}\right)^{n+1}
\,.
\end{align}

\begin{figure}[t]
\begin{center}
\includegraphics[width=11cm]{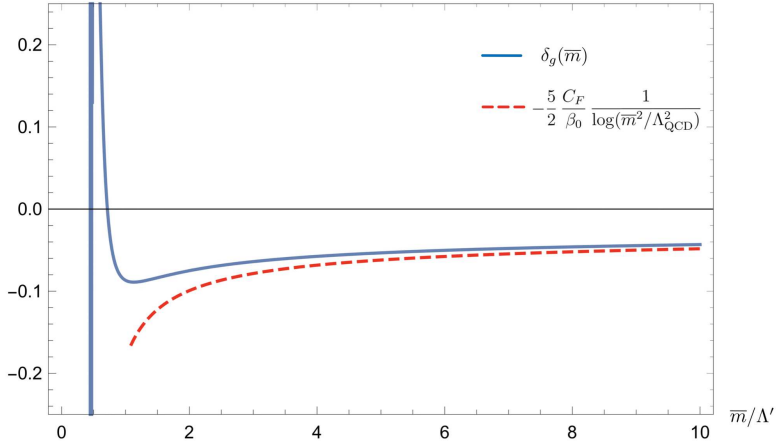}
%\includegraphics[width=8cm]{Fig-delta_g.eps}
%\vspace*{-.5cm}
\caption{\footnotesize
$\delta_g$ vs.\ $\mbar/\Lambda'$ in the case that
the number of light quark flavors is zero, i.e., $\beta_0=11$.
Dashed line shows the leading asymptotic term for large $\mbar$, 
eq.~\eqref{asympt-deltag}.
\label{Fig-delta_g}}
\end{center}
\end{figure}

Thus, the above series is naturally separated into two parts:
\begin{align}
&
\delta = \delta_G + \delta_g 
\,,
\\
&
\delta_G(\mu=\mbar)=
\frac{C_F\alpha_s(\mbar)}{2\pi}\sum_{n=0}^\infty\left(\frac{\beta_0\alpha_s(\mbar)}{4\pi}\right)^n
\,G_{n+1}(\mbar)\cdot n!
\,,
\label{series-deltaG}
\\
&
\delta_g(\mu=\mbar)=
\frac{C_F\alpha_s(\mbar)}{2\pi}\sum_{n=0}^\infty\left(\frac{\beta_0\alpha_s(\mbar)}{4\pi}\right)^n
\frac{(-1)^n}{n+1}g_{n+1}
\,.
\end{align}
Here and hereafter, we set $\mu=\overline{m}$.
[We are interested in the difference of
the RG invariant masses
$\delta m(\mbar)= m_{\rm pole}-\mbar (\mbar)$.]
Eq.~\eqref{gn-anomalousdim}
shows that $g_n$'s are determined completely by the mass anomalous dimension
within the large-$\beta_0$ approximation.
Namely, $g_n$'s are determined by the UV divergences relevant to 
the heavy quark {\msbar} mass renormalization.
Hence, it is natural to consider $\delta_g$ as a genuinely UV quantity.
This series expansion has a non-zero (finite) 
radius of convergence about $\alpha_s(\mbar)=0$
and can be expressed by $\gamma_m^{(\beta_0)} $ 
through 
eq.~\eqref{gn-anomalousdim} as
\begin{align}
\delta_g
=
-\frac{2C_F}{\beta_0}
\int_0^{-a}
\!
dx\, \frac{g(x)-g_0}{x}
~~~~~~;~~~~~~~
a=\frac{\beta_0\alpha_s(\mbar)}{4\pi}
\,.
\end{align}
In Fig.~\ref{Fig-delta_g} we plot $\delta_g$ as a function of $\mbar$.
The leading behavior of $\delta_g$ for large $\mbar$
is given by
\begin{align}
\delta_g (\mbar)
\to \frac{C_F\alpha_s(\mbar)}{2\pi}\, g_1 = -\frac{5}{2}\,\frac{C_F}{\beta_0}\,
\frac{1}{\log(\mbar^2/\LQ^2)}
~~~\mbox{for}~~~
\mbar \gg \LQ 
\label{asympt-deltag}
\,.
\end{align}
$\delta_g$ becomes highly oscillatory at $\mbar/\Lambda' \lesssim 1$ 
reflecting the oscillatory behavior of $g(t)$ at $t \lesssim -1$.

On the other hand,
$\delta_G$ includes IR renormalons, hence the series is asymptotic
(convergence radius is zero).
We can separate a genuine UV part and IR sensitive part of $\delta_G$ by
the CD prescription.
By appropriately subtracting the UV divergence, we can write
\begin{align}
\delta_G (\mbar)
&
= \lim_{M\to\infty}\left[
\int ^{M^2 }_{0}\frac {d\tau }{\pi \tau }\,
{\rm Im}\,\overline{W}_m\!\left(\dfrac{\tau}{\overline{m}^2}\right)
\,
{\alpha_{\beta_0}(\tau)} - f_{\rm UV}
\right]
\label{deltaG1} 
\\ 
&=
\int ^{\infty}_{0}\frac {d\tau }{\pi \tau }\,
\left[
{\rm Im}\,\overline{W}_m\!\left(\dfrac{\tau}{\overline{m}^2}\right)
-\frac{3C_F}{4\pi}\, \theta (\tau-e^{5/3}\mbar^2)
\right]
{\alpha_{\beta_0}(\tau)}
\,,
\label{deltaG2}
\end{align}
where\footnote{
Roughly speaking,
to obtain only the $G$ part, we can take the $\epsilon^0$-part
before $\tau$-integration.
}
\begin{align}
\overline{W}_m(s)=
-\frac {C_{F}}{4\pi }\int ^{1}_{0}dx\,
\left[(2+2x) \log\left( {x^{2}+s\, x-s -i0}\right)+2x\right]
\,,
\end{align}
and
\begin{align}
f_{\rm UV}=\frac{3C_F }{\beta_0}
\log \left[ \frac{\log(M^2/\Lambda'^2)}{\log(\overline{m}^2/\LQ^2)}
\right]
\,.
\end{align}
In the equality of eq.~\eqref{deltaG2} we have expressed
$f_{\rm UV}$ in an integral form and taken the limit $M\to\infty$;
$\theta(x)$ denotes the unit step function.
$\overline{W}_m(s)$ can be expressed in terms of elementary functions.
By expanding eq.~\eqref{deltaG1} or 
\eqref{deltaG2} in $\alpha_s(\mbar)$ one can  check
that eq.~\eqref{series-deltaG} is reproduced.

Now we can apply the CD prescription to the first term of eq.~\eqref{deltaG1}. 
We introduce a factorization scale $\mu_f (\gg \Lambda')$ and
restrict the integral region as $(0,M^2) \to (\mu_f^2,M^2)$.
Then the integral becomes well defined 
%after resummation to all orders of $\alpha_s(\mbar)$,
avoiding the singularity of $\alpha_{\beta_0}(\tau)$.
We can separate the integral to a genuine UV part (independent of $\mu_f$)
and IR sensitive part (dependent on $\mu_f$).
Using the expansion of $\overline{W}_m$ for $|\tau/\mbar^2| \ll 1$,
\begin{align}
\overline{W}_m=\frac {C_{F}}{4\pi }
\left[
4+2\pi i \left(\frac{\tau}{\mbar^2}\right)^{1/2} - 3\left(\frac{\tau}{\mbar^2}\right)
-\frac{3\pi i}{4}\left(\frac{\tau}{\mbar^2}\right)^{3/2}  + \cdots
\right]
\,,
\end{align}
we obtain
\begin{align}
\delta_{G,UV}
&
\equiv \lim_{M\to\infty}\left[
{\rm Im}\int ^{M^2 }_{\mu_f^2}\frac {d\tau }{\pi \tau }\,
\overline{W}_m\!\left(\dfrac{\tau}{\overline{m}^2}\right)
\,
{\alpha_{\beta_0}(\tau)} - f_{\rm UV}
\right]
\nonumber \\
&
= \delta_{G,0}(\mbar)
 + \frac{C_0^m(\mu_f)}{\mbar} - \frac{3C_F}{\beta_0} \frac{\Lambda'^2}{\mbar^2}
+{\cal O}(\LQ^3/\mbar^3)
\,.
\label{deltaG-UV}
\end{align}

\begin{figure}[t]
\begin{center}
\includegraphics[width=6cm]{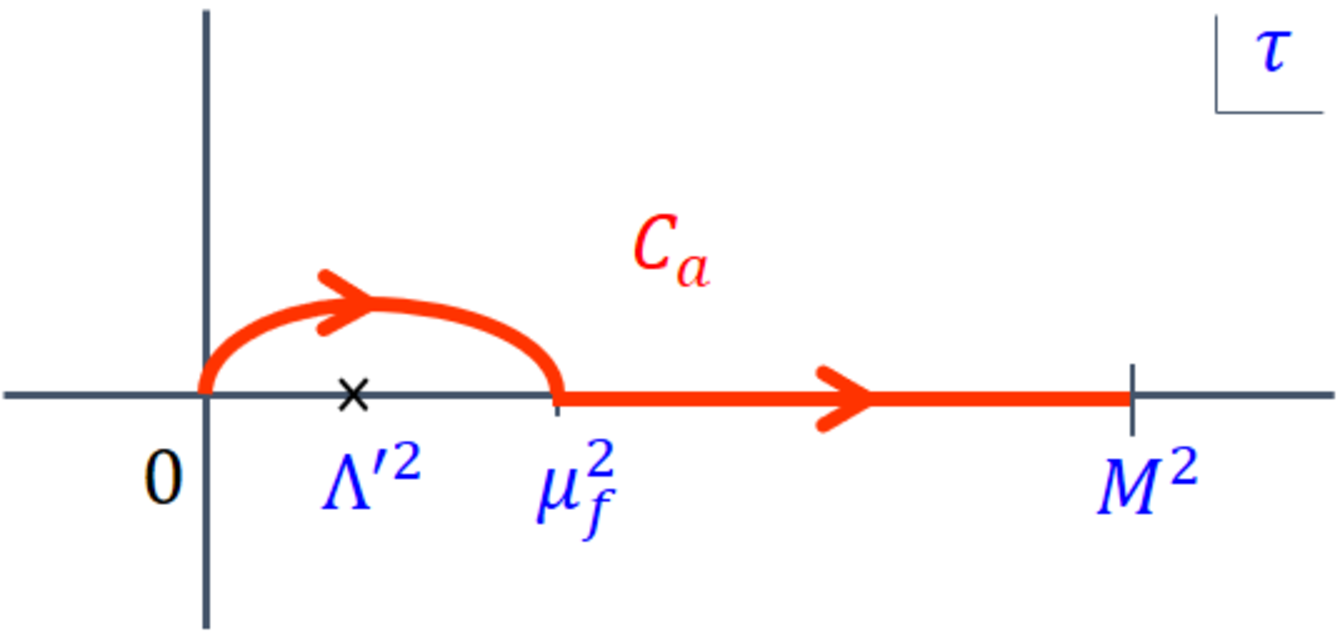}
~~~~~~~
\includegraphics[width=6cm]{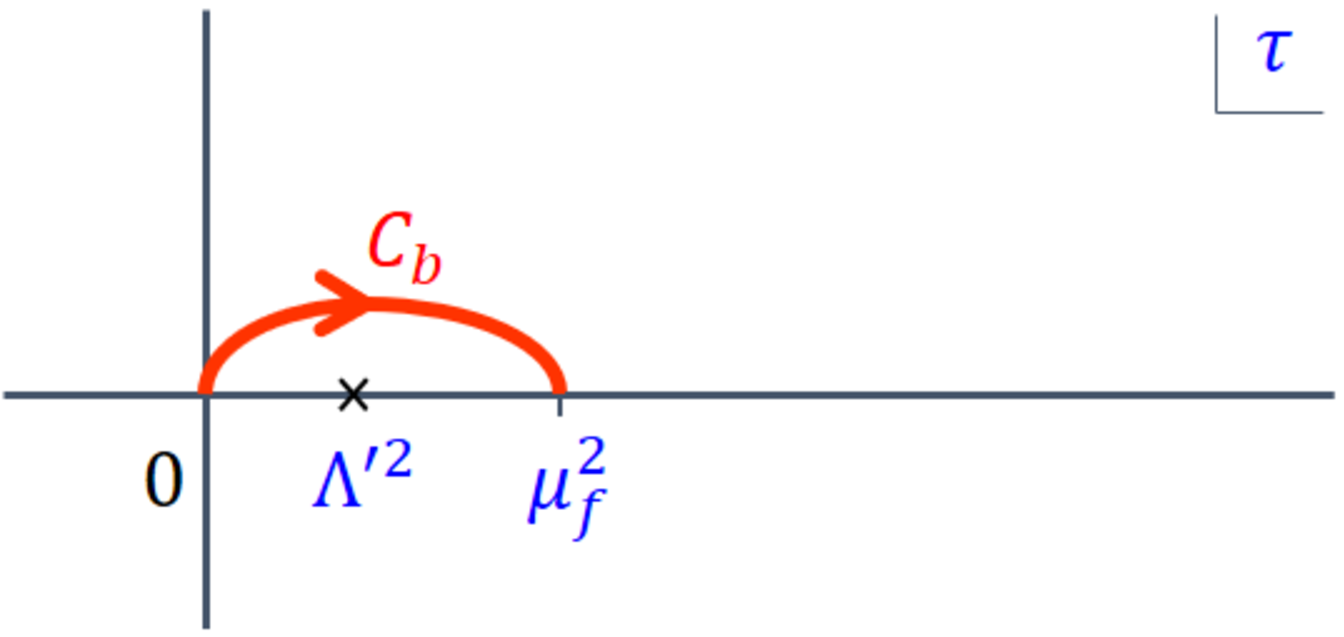}
\\
(a)\hspace{70mm}(b)
\vspace*{-3mm}
\end{center}
\caption{\footnotesize
Integral contours in the complex $\tau$-plane shown by
red lines.
The pole position of $\alpha_{\beta_0}(\tau)$ is also shown.
\label{Contours}}
\end{figure}

\begin{figure}[t]
\begin{center}
\includegraphics[width=11cm]{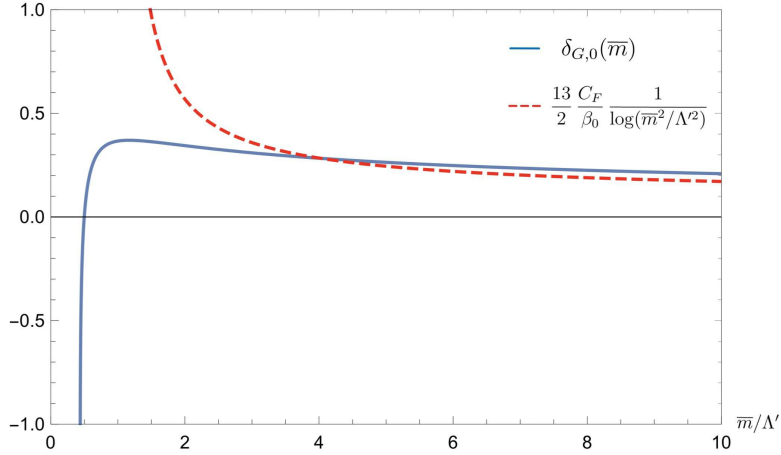}
%\includegraphics[width=8cm]{Fig-delta_G0.eps}
%\vspace*{-.5cm}
\caption{\footnotesize
$\delta_{G,0}$ vs.\ $\mbar/\Lambda'$ in the case that
the number of light quark flavors is zero.
Dashed line shows the leading asymptotic term for large $\mbar$, 
eq.~\eqref{asympt-deltaG0}.
%$\delta_{G,0}$ approaches its asymptotic form much more slowly than
%$\delta_{g}$.
\label{Fig-delta_G0}}
\end{center}
\end{figure}

The first and third terms of the
expansion are independent of $\mu_f$.
The first term is given by
\begin{align}
\delta_{G,0}(\mbar)
&=
\lim_{M\to\infty}\left[
\frac{4C_F}{\beta_0}+
{\rm Im}\int_{C_a}\frac {d\tau }{\pi \tau }\,
\overline{W}_m\!\left(\dfrac{\tau}{\overline{m}^2}\right)
\,
{\alpha_{\beta_0}(\tau)} - f_{\rm UV}
\right]
\label{delta_G,0}\\
&=
\frac{C_{F}}{\pi\beta_{0}}\int_{0}^{\infty} ds\, w(s)\, A\left(s\mbar^2/\Lambda'^2\right)+\frac{3 C_{F}}{\beta_{0}} \log \left[\frac{\log \left(\overline{m}^{2} / \Lambda_{\rm{QCD}}^{2}\right)}{\pi/e}\right]
\,,
\label{delta_G,0-improved}
\end{align}
where
\begin{align}
A(x)=\frac{\pi}{2}-\arctan \left[\frac{\log (x)}{\pi}\right]
\,,
\end{align}
and
\begin{align}
&
w(s)=-2+s \log s-2\left(s^{2}-2 s-2\right)T(s)-\frac{3}{s}\,\theta\left(s-\frac{\Lambda'^{2}}{\overline{m}^{2}}\right)
\,,
\\&
T(s)=\theta (4-s)
\frac{\arctan \left(\sqrt{\frac{4-s}{s}}\right)}{\sqrt{s(4-s)}}+
\theta(s-4)
\frac{{\rm arctanh} \left(\sqrt{\frac{s-4}{s}}\right)}{\sqrt{s(s-4)}} \,.
%T(s)=\left\{
%\begin{array}{l}
%\frac{\arctan \left[\sqrt{\frac{4-s}{s}}\right]}{\sqrt{s(4-s)}} \,, ~~(s<4)
%\\
%\frac{{\rm arctanh} \left[\sqrt{\frac{s-4}{s}}\right]}{\sqrt{s(s-4)}} \,, ~~(s>4)
%\end{array}
%\right.
\end{align}
The integral contour $C_a$ is shown in Fig.~\ref{Contours}(a).
In eq.~\eqref{delta_G,0-improved}, we rotate the contour to the negative real $\tau$-axis 
($\tau=e^{i\pi}\, s \, \overline{m}^{2} / \Lambda'^{2}$)
in order to remove the $M^2$ dependence between the second and third terms of eq.~\eqref{delta_G,0}.
The dependence of $\delta_{G,0}(\mbar)$ on $\mbar/\Lambda'$ is shown in Fig.~\ref{Fig-delta_G0}.
Its asymptotic form is given by
\begin{align}
\delta_{G,0}(\mbar)
& \to \frac{13}{2}\, \frac{C_F}{\beta_0}\,\frac{1}{\log(\mbar^2/\Lambda'^2)}
~~~\mbox{for}~~~
\mbar/\Lambda' \gg 1
\label{asympt-deltaG0}
\,.
\end{align}

The second term of eq.~\eqref{deltaG-UV} depends on $\mu_f$. 
If we multiply the term by $\mbar$, it is independent of 
$\mbar$ and is given by
\begin{align}
C_0^m(\mu_f)=-\frac{C_F}{2\pi}\,{\rm Re}\int_{C_b}
\frac{d\tau}{\sqrt{\tau}} \, \alpha_{\beta_0}(\tau) 
\,,
\end{align}
where the integral contour $C_b$ is shown in Fig.~~\ref{Contours}(b).
The corresponding term in the QCD potential has exactly the form such that
$2 C_0^m(\mu_f) + C_0^V(\mu_f)=0$ in the CD prescription \cite{Sumino:2004ht}, 
showing the cancellation of renormalons in the
$r$ and $\mbar$-independent part of $E_{\rm tot}(r)$; c.f., eqs.~\eqref{expVQCD}
and \eqref{expmpole}.

Thus, $\Delta M_0(\mbar)$ in eq.~\eqref{expmpole} is given by
\begin{align}
\Delta M_0(\mbar)= (\delta_g + \delta_{G,0}) \cdot \mbar
\,.
\end{align}
%Its dependence on $\mbar$ is shown in Fig.~.
The asymptotic form is determined by the sum of eqs.~\eqref{asympt-deltag} and 
\eqref{asympt-deltaG0} and reads 
\begin{align}
\Delta M_0(\mbar)
& \to \, \frac{4C_F}{\beta_0}\,\frac{\mbar}{\log(\mbar^2/\LQ^2)}
~~~\mbox{for}~~~
\mbar/\LQ \gg 1
\,.
\end{align}
It agrees with the requirement by RG for the leading asymptotic behavior
of the mass difference $m_{\rm pole}-\mbar(\mbar)$.

As can be seen from Figs.~\ref{Fig-delta_g} and \ref{Fig-delta_G0}
(see also Fig.~\ref{Fig-mdep-part-Etot} below), the
behavior of $\Delta M_0(\mbar)$ at $\mbar \gtrsim \Lambda'$ is
consistent with the expectation that it is proportional to $\mbar$
with logarithmic corrections at large $\mbar$.
At small $\mbar(\lesssim \Lambda'$), however, $\Delta M_0(\mbar)$ has an oscillatory
behavior.
This feature is absent in the corresponding expansions of 
the static potential $V_{\rm QCD}(r)$ and the Adler function
\cite{Sumino:2005cq,Mishima:2016xuj}, whose radiative corrections 
are dominated by those in Euclidean
regions.
The oscillatory behavior
may reflect the fact that in the pole mass the
self-energy corrections close to the on-shell quark configuration
involve time-like kinematics.
In any case, since this behavior can be concealed by 
${\cal O}(\LQ^3/\overline{m}^2)$ renormalon uncertainties,
we cannot make any definite statement about it.

The final result for the genuinely UV (renormalon-free) part of the total energy is given by
\begin{align}
E_{\rm tot}(r) =
V_C(r) +  C_1^V  r +
2\left[
\overline{m} + \Delta M_0(\overline{m}) 
+  \frac{C_1^m}{\overline{m}} 
\right]
+ {\cal O}(\LQ^3 r^2, \, \LQ^3/ \overline{m}^2) 
\,.
\label{final-result}
\end{align}
with
\begin{align}
&
C_1^m  = -\frac{3 C_F}{\beta_0} \Lambda'^2
\,.
\end{align}
The $\mbar$-dependent part, 
$\Delta M_0(\overline{m}) 
+  {C_1^m}/{\overline{m}} $,
is shown in Fig.~\ref{Fig-mdep-part-Etot} 
as a function of $\mbar/\Lambda'$.
We see that the contribution of the $C_1^m/\mbar$ term quickly
diminishes at $\mbar/\Lambda' \gtrsim 2$.
For completeness we also list the known results for $V_C(r)$ and $C_1^V$
in the large-$\beta_0$ approximation:
\begin{align}
&
V_C(r)=-\frac{4C_F}{\beta_0r}
\int_0^\infty\!\! dt \, {e^{-t}}\, A\left(t^2/(\Lambda'^2r^2)\right)\,,
\\
&
C_1^V  = \frac{2 \pi C_F}{\beta_0} \Lambda'^2
\,.
\end{align}

\begin{figure}[t]
\begin{center}
\includegraphics[width=13cm]{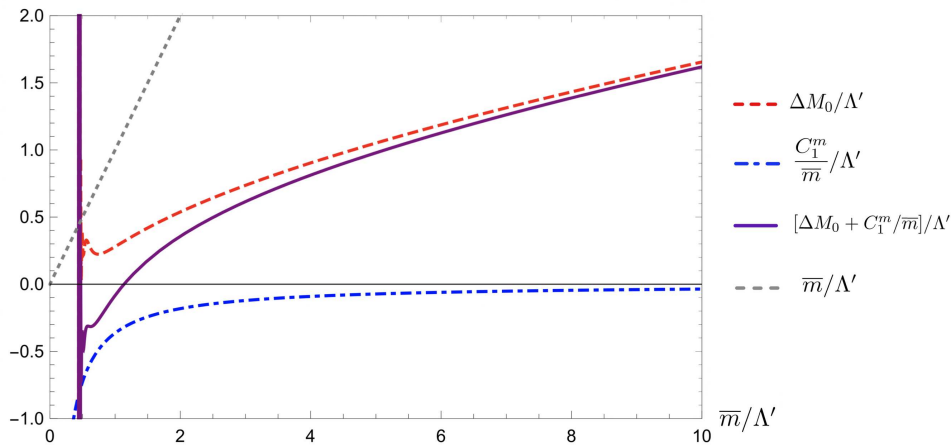}
%\includegraphics[width=8cm]{Fig-mdep-part-Etot.eps}
%\vspace*{-.5cm}
\caption{\footnotesize
$[\Delta M_0 +  {C_1^m}/{\overline{m}}]/\Lambda' $ 
vs.\ $\mbar/\Lambda'$ in the case that
the number of light quark flavors is zero  (purple solid line).
Individual terms are shown in red dashed 
($\Delta M_0/\Lambda' $)
and blue dot-dashed (${C_1^m}/({\overline{m}}\Lambda') $) lines.
For comparison, $\mbar/\Lambda'$ is shown as a gray dotted line.
\label{Fig-mdep-part-Etot}}
\end{center}
\end{figure}

Let us present some speculation.
$\Delta M_0(\overline{m})$ consists of $\mbar \cdot \delta_g$, the part which can be
expanded in the Taylor series in $\alpha_s(\mbar)$, and 
$\mbar \cdot \delta_{G,0}$, the part which cannot be expanded
in the Taylor series in $\alpha_s(\mbar)$
[expansion in $1/\log(\mbar/\LQ)$ is asymptotic].
%For large $\mbar$, $\delta_g$ is expanded in powers of 
%$\alpha_s(\mbar) \propto 1/\log(\mbar/\LQ)$, whereas
%$\delta_{G,0}$ is more naturally regarded as the leading term
%of the power series expansion in $\LQ/\mbar$
%(with logarithmic corrections).
The former is tightly connected with the renormalization of the {\msbar} mass
and is intrinsic to this mass scheme.
The latter originates from the UV part of the asymptotic series 
$\sim G_{n+1}\cdot n!$, and hence is likely to be tied to the pole
mass, irrespective of the definition of the short-distance mass.
%Both contribute\footnote{
%This feature can be seen from the series expansion eq.~\eqref{all-order-formula},
%where both $G_1$ and $g_1$ terms contribute to the leading
%asymptotic form.
%}
%to the asymptotic behavior of $m_{\rm pole}-\mbar(\mbar)$
%for large $\mbar$.

There is no $r$-independent or $\mbar$-independent term proportional to
$\LQ$ in the latter part of $E_{\rm tot}(r)$
in expansions in $r$ and $1/\mbar$.
This would be natural with regard to the fact that  the size of a heavy
quarkonium bound state is small compared to ordinary hadrons.
As a whole we consider that the expression of 
$E_{\rm tot}(r)$ in eq.~\eqref{final-result} has a natural form:
It includes only UV contributions;
It is composed of the part
which conforms with power expansions in
$1/\mbar$ and $r$ ($\delta_{G,0}$, $C_1^m$, $V_C$ and $C_1^V$) 
and the part which can be expanded in the Taylor series in powers of  
$\alpha_s(\mbar)\sim 1/\log (\mbar/\LQ)$
whose expansion coefficients
originate from the UV divergences of the short-distance mass
($\delta_g$).\footnote{
Noting that $\mbar$ is much larger than $\LQ$, in 
numerical analyses it is not trivial to identify an order $\LQ$ constant
in $\Delta M_0(\overline{m})$ which is of order $\mbar/\log\mbar$.
Thus, we see an advantage of performing a semi-analytic analysis. 
}
In turn, this can be taken
as an evidence that we have chosen a sensible scheme
for separating the UV and IR contributions and performing expansions
in $r$ and $1/\mbar$.\footnote{
The CD prescription is known to have a scheme dependence.
The standard scheme choice (``massive gluon scheme'') is 
favorable from the viewpoint of analyticity \cite{Mishima:2016vna},
and we have chosen this scheme in the above computation.
}

%The oscillatory behavior of $\delta_g$ in the region $\mbar/\Lambda' \lesssim 1$
%may reflect the fact that the
%self-energy corrections close to the on-shell quark configuration
%involve time-like kinematics.
%In any case, since this behavior can be concealed by 
%${\cal O}(\LQ^3/\overline{m}^2)$ renormalon uncertainties,
%we cannot make any definite prediction about it.

Finally we comment on possible existence of ${\cal O}(\LQ^2/m)$  
renormalon contained 
in the pole mass, whose properties are as yet not well known.
Known properties are as follows \cite{Neubert:1996zy}.
(a) It is induced by the non-relativistic kinetic energy
operator $\vec{D}^2/(2m)$;
(b) It is not forbidden by any symmetry, and corresponds
to the singularity at $u=1$ in the Borel plane;
(c) It does not appear in the large-$\beta_0$ approximation.
Thus, it could affect the computable part of $E_{\rm tot}(r)$
beyond the large-$\beta_0$ approximation.
We leave this issue to future study.\footnote{
This renormalon may eventually cancel out due to off-shell effects
\cite{Kiyo:2001zm}.
}

\section*{Acknowledgements}

The authors are grateful to fruitful discussion with H.~Takaura.
The work of Y.S.\
was supported in part by Grant-in-Aid for
scientific research (No.\  17K05404) from
MEXT, Japan.
%The works of Y.H.\ and Y.S., respectively,
%were supported in part by Graduate Program on Physics for
%the Universe (GP-PU), Tohoku University, and by
%Grant-in-Aid for
%scientific research (No.\  17K05404) from
%MEXT, Japan.

\end{document}